\documentclass[12pt]{amsart}
\usepackage{amsmath,amssymb,amsthm}
\usepackage{color}
\usepackage{hyperref}

\newtheorem{lem}{Lemma}[section]
\newtheorem{prop}{Proposition}[section]

\newtheorem{thm}{Theorem}[section]

\theoremstyle{definition}

\theoremstyle{remark}

\theoremstyle{remark}
\newtheorem{remark}{Remark}[section]
\numberwithin{equation}{section}

\newcommand{\N}{{\mathbb N}}

\newcommand{\R}{{\mathbb R}}

\definecolor{blu}{rgb}{0,0,1}

\begin{document}

\title[Ground states for semi-relativistic  SPS energy]{Ground states for semi-relativistic  
Schr\"odinger-Poisson-Slater energy }
\author{Jacopo Bellazzini}
\address{J. Bellazzini, 
\newline  Universit\`a di Sassari, Via Piandanna 4, 07100 Sassari, Italy}%
\email{jbellazzini@uniss.it}%
\author{Tohru Ozawa}
\address{T. Ozawa
\newline Department of Applied Physics, Waseda University, Tokyo 169-8555, Japan}%
\email{txozawa@waseda.jp}%
\author{Nicola Visciglia}
\address{N. Visciglia, \newline Dipartimento di Matematica Universit\`a di Pisa
Largo B. Pontecorvo 5, 56100 Pisa, Italy}%
\email{viscigli@dm.unipi.it}

\maketitle

\begin{abstract}
We prove the existence of ground states for the semi-relativistic  
Schr\"odinger-Poisson-Slater energy
$$I^{\alpha,\beta}(\rho)=\inf_{\substack{u\in H^\frac 12(\R^3)\\\int_{\R^3}|u|^2 dx=\rho}} 
\frac{1}{2}\|u\|^2_{H^\frac 12(\R^3)} 
+\alpha\int\int_{\R^{3}\times\R^{3}}
\frac{| u(x)|^{2}|u(y)|^2}{|x-y|}dxdy-\beta\int_{\R^{3}}|u|^{\frac{8}{3}}dx$$
 $\alpha,\beta>0$ and  
$\rho>0$ is small enough. 
The minimization problem is  $L^2$ critical and  in order to characterize of the values $\alpha, \beta>0$ such that 
$I^{\alpha, \beta}(\rho)>-\infty$ for every $\rho>0$, we prove a new lower bound on the Coulomb energy involving the kinetic energy and the exchange energy. We prove the existence of a constant $S>0$ such that
$$\frac{1}{S}\frac{\|\varphi\|_{L^\frac 83(\R^3)}}{\|\varphi\|_{\dot H^\frac 12(\R^3)}^\frac 12}\leq 
\left (
\int\int_{\R^3\times \R^3}
\frac{|\varphi(x)|^2|\varphi(y)|^2}{|x-y|}dxdy\right )^\frac 18
$$
for all $\varphi\in C^\infty_0(\R^3)$.
Eventually we show that similar compactness property fails  provided that
in the energy above we replace the inhomogeneous Sobolev norm
$\|u\|^2_{H^\frac 12(\R^3)}$ by the homogeneous one $\|u\|_{\dot H^\frac 12(\R^3)}$.
\end{abstract}

Aim of this paper is to prove the existence of ground states for 
the following minimization problem:
\begin{equation}\label{minim}
I^{\alpha,\beta}(\rho)=\inf_{u\in S(\rho)} {\mathcal E}^{\alpha,\beta}(u)
\end{equation}
where
\begin{equation}\label{II}
{\mathcal E}^{\alpha,\beta}(u)=\frac{1}{2}\|u\|^2_{H^\frac 12(\R^3)} 
+\alpha\int\int_{\R^{3}\times\R^{3}}
\frac{| u(x)|^{2}|u(y)|^2}{|x-y|}dxdy- \beta\int_{\R^{3}}|u|^{\frac{8}{3}}dx,
\end{equation}
$$\alpha, \beta>0$$
\begin{equation}\label{srho}
S(\rho)=\left \{u\in H^{\frac 12}(\R^{3}) \hbox{ s.t. } \int_{\R^3} |u|^2 dx =\rho\right \}
\end{equation}
and $H^s (\R^3)$ denotes for general $s\in \R$ the usual
Sobolev spaces endowed with the norm:
$$\|u\|_{H^s(\R^3)}^2= \int_{\R^3} (1+|\xi|^2)^s |\hat u(\xi)|^2 d\xi$$
with 
$\hat u(\xi)=\int_{\R^3} e^{{-2\pi \bf i}x\dot \xi} u(x) dx.$\\
\\
The aforementioned minimization problem  arises from statistical physics, being the semi-relativistic version of the  Hartree-Fock energy proposed by Slater \cite{Sl}
for a system of electrons
interacting with each other via the Coulomb law. In the  Hartree-Fock model proposed by Slater \cite{Sl} the focusing term
$||u||_{\frac{8}{3}}^{\frac{8}{3}}$ is the exchange energy due the Pauli principle   and $\iint{| u(x)|^{2}|u(y)|^2}{|x-y|}dxdy$
describes the repulsive Coulomb interaction. The quantity $\rho$ measures  the total number of electrons.\\
\\
In this paper we treat the semi-relativistic case, i.e considering the kinetic term given by $\|u\|^2_{H^\frac 12(\R^3)}$ instead of the classical term $\|u\|^2_{H^1(\R^3)}$ proposed by Slater  \cite{Sl}. For this reason we call \eqref{II} the semi-relativistic Schr\"odinger-Poisson-Slater energy. The main question addressed in this paper is the role of $\alpha, \beta$ and $\rho$ in the minimization problem.
\\
It is important to underline from the beginning the main difficulties to prove the existence of minimizers for the semi-relativistic Schr\"odinger-Poisson-Slater energy:
\begin{itemize}
\item the problem is $L^2$ critical, namely it is not sufficient to apply  the fractional Gagliardo-Niremberg inequality to get $I^{\alpha,\beta}(\rho)>-\infty $ for all $\rho$
\item it is not clear if one can choose a sequence of \emph{radially symmetric} functions as a minimizing sequence due to the competition between the Coulomb term and the kinetic  energy
\item in $H^{\frac{1}{2}}(\R^3)$ without symmetry informations it not straightforward to prove that a bounded minimizing sequence with some additional assumption has a non vanishing weak limit
\item it is not elementary to avoid \emph{dichotomy}, i.e to prove that the weak limit belongs to $S(\rho)$, due to the presence of three terms in the energy functional
\end{itemize}

Recall that a general strategy to attack constrained minimization problems is the celebrated 
concentration-compactness principle of P.L. Lions, see \cite{L}. 
The main point is that in general if $u_n$ is a bounded minimizing sequence
for \eqref{minim} then  
up to translations two possible bad scenarios can occur (that
can be shortly summarized as follows):
\begin{itemize}
\item (vanishing) $u_n\rightharpoonup 0$;
\item (dichotomy) $u_n \rightharpoonup \bar u\neq 0$ and $0<\|\bar u\|_2<\rho$.
\end{itemize}
Typically the vanishing can be excluded by proving that 
any minimizing sequence weakly converges, 
up to translation, 
to a function $\bar u$ different from zero (in turn it can be accomplished in general
by a suitable localized Gagliardo-Nirenberg inequality
in conjunction with the Rellich compactness theorem).\\
Concerning the dichotomy the classical way to rule out it 
is by proving the following strong subadditivity inequality 
\begin{equation}\label{intro1}
I^{\alpha,\beta}(\rho)< I^{\alpha,\beta}(\mu)+I^{\alpha,\beta}(\rho-\mu) 
\  \text{ }\ \forall  \hbox{ } 0<\mu<\rho.
\end{equation}
Although
the following weak version 
of \eqref{intro1}
\begin{equation}\label{intro2}
I^{\alpha,\beta}(\rho) \leq I^{\alpha,\beta}(\mu)+
I^{\alpha,\beta}(\rho-\mu) \ \ \text{ for all }\  0<\mu<\rho.
\end{equation}
can be easily proved, in general the proof of \eqref{intro1}
requires some extra arguments which heavely depend on the structure
of the functional we are looking at.\\

The existence of minimizers for semi-relativistic energies is not a novelty, see e.g  
\cite{LL}, \cite{LY}, \cite{FJL} for the case $\beta=0$ and $\alpha<0$ .  We shall underline however that when $\beta=0$ and $\alpha<0$, the Boson star minimization problem, a sequence of radially symmetric functions can be chosen
as minimizing sequence.  On the other hand when $\alpha>0$ and $\beta>0$ the only known results concern the existence of ground states 
for the classical  Schr\"odinger-Poisson-Slater 
energy (i.e. \eqref{minim} where 
$||u||_{H^{\frac{1}{2}}}^2$  is replaced by  $||u||_{H^{1}}^2$).  In the classical case the existence of minimizers for small $\rho$ is proved in
\cite{SS}   in case $\alpha, \beta>0$,
and extended in \cite{BS} and \cite{BS2} 
if one replaces the exponent $\frac{8}{3}$ respectively with $3<p<\frac{10}{3}$ and  $2<p<3$ (see also \cite{CDSS} for a review paper on the subject). Finally we quote \cite{CL} where it is studied the non relativistic 
Schr\"odinger-Poisson-Slater equation with the nonlinearity 
$|u|^{\frac{10}{3}}-|u|^{\frac{8}{3}}$. Notice that for the classical  Schr\"odinger-Poisson-Slater the minimization problem is $L^2$ sub-critical, namely it is straighforward to show
that the energy is bounded from below and that the minimizing sequence is bounded.\\
Now we are ready to state our main results.  The first result concerns the characterization of the values
$\alpha, \beta>0$ such that 
$I^{\alpha, \beta}(\rho)>-\infty$ for every $\rho>0$.\\
We need to introduce
the constant $S$ defined as follows:
$$S=\inf \{C\in (0, \infty] \hbox{ s.t. } C \hbox{ satisfies }\eqref{impaz}\}$$
\begin{equation}\label{impaz}\|\varphi\|_{L^\frac 83(\R^3)}
\leq C \|\varphi\|_{\dot H^\frac 12(\R^3)}^\frac 12
\left (\int\int_{\R^3\times \R^3} 
\frac{|\varphi(x)|^2|\varphi(y)|^2}{|x-y|}dxdy\right )^\frac 18 
\end{equation}
$$
\forall \varphi\in C^\infty_0(\R^3)
$$
where
$\|\varphi\|_{\dot H^\frac 12 (\R^3)}^2= \int_{\R^3} |\xi| |\hat \varphi(\xi)|^2 d\xi.$
In the next section we prove that the estimate \eqref{impaz}
is true and hence $S<\infty$ is its best constant.
\begin{thm}\label{bestc} Let $\alpha, \beta>0$ be fixed.
Then the following facts are equivalent: 
\begin{itemize}\item $\exists \rho>0 \hbox{ s.t. } 
I^{\alpha, \beta}(\rho)=-\infty;$
\item $\left(\frac{27\alpha}{\beta^3}\right )^\frac 18< \sqrt 2S.$
\end{itemize}
\end{thm}
The second result concerns the existence of minimizers for   the semi-relativistic Schr\"odinger-Poisson-Slater energy.
\begin{thm}\label{XXXX}
For every $\alpha, \beta>0$ there exists $\bar \rho=\bar \rho(\alpha, \beta)>0$ 
such that $I^{\alpha, \beta}(\rho)\geq 0$ for every $0<\rho<\bar \rho$.
Moreover for every sequence $u_{n}$ which satisfy: 
$$u_{n}\in S(\rho) \hbox{ and }  {\mathcal E}^{\alpha,\beta}(u_n)\rightarrow
I^{\alpha,\beta} (\rho), 
\hbox{ with } 0<\rho<\bar \rho$$
there exists, up to subsequence, $\tau_n\in \R^3$ such that
$$u_{n}(.+\tau_n) \hbox{ has a strong limit in } H^\frac 12(\R^3).$$
In particular the set of minimizers for $I^{\alpha,\beta}(\rho)$ is not empty
for $\rho$ small.
\end{thm}
In our opinion theorem \ref{XXXX} 
is quite surprising in view of the next nonexistence result.
First we introduce the following 
minimization problems
$$\tilde I^{\alpha,\beta}(\rho)=\inf_{u\in S(\rho)} \tilde {\mathcal E}^{\alpha,\beta}(u)
$$
where
\begin{equation}\label{IItilde}
\tilde {\mathcal E}^{\alpha,\beta}(u)=\frac{1}{2}\|u\|^2_{\dot H\frac 12(\R^3)} 
+\alpha\int\int_{\R^{3}\times\R^{3}}
\frac{| u(x)|^{2}|u(y)|^2}{|x-y|}dxdy-\beta\int_{\R^{3}}|u|^{\frac{8}{3}}dx.
\end{equation}
and
$$\|u\|_{\dot H^\frac 12 (\R^3)}^2= \int_{\R^3} |\xi| |\hat u(\xi)|^2 d\xi$$
(here $\hat u(\xi)$ denotes the Fourier transform of $u$
and $S(\rho)$ is defined in \eqref{srho}).
Notice that the unique difference between 
${\mathcal E}^{\alpha,\beta}$ and $\tilde {\mathcal E}^{\alpha,\beta}$ concerns the 
quadratic part which in the second case is an homogeneous norm, while in the first case is
the inhomogeneous one.
\begin{thm}\label{homogeneous}
For every $\alpha, \beta>0$ there exists $\bar \rho=\bar \rho(\alpha, \beta)>0$ such that:
\begin{itemize}
 \item 
$\tilde I^{\alpha,\beta}(\rho)>-\infty \hbox{ } \forall \rho\in (0, \bar \rho);$
\item $\forall \rho\in (0, \bar \rho) \hbox{ and } \forall v\in S(\rho) \hbox{ we have }$
$\tilde {\mathcal E}^{\alpha,\beta}(v)>\tilde I^{\alpha,\beta}(\rho)$\\
(i.e. there are not minimizers for $\tilde I^{\alpha,\beta}(\rho)$ with $\rho$ small).
\end{itemize}
\end{thm}
$$$$
\begin{remark}
The statement of Theorem \ref{XXXX} does not change if the exchange energy is replaced by $||u||_p^p$ for $2<p<\frac{8}{3}$. In this case
the minimization problem is $L^2$ subcritical such that the energy is bounded from below for all $\alpha, \beta>0$. The existence of ground states   $2<p<\frac{8}{3}$ follows 
as in Theorem \ref{XXXX}.
\end{remark}
We conclude the introduction discussing the connection between minimizer for for \eqref{minim} and steady states of a suitable semi-relativistic 
nonlinear Schr\"odinger Equation.\\
By using the well--known property $\||w|\|^2_{H^\frac 12(\R^3)}\leq \|w\|^2_{H^\frac 12(\R^3)}$,
where equality occurs if and only if there exists $\theta\in \R$ such that
$e^{i\theta}w$ is real--valued (see for instance \cite{LLo}),
one can deduce that if $v(x)$ is a minimizer for \eqref{minim}
then there exists $\theta\in \R$ such that
$e^{{\bf i}\theta}v$ is real--valued. In particular any minimizer 
$v$ to \eqref{minim} solves the following equation:
\begin{equation}\label{eq}
\sqrt{1-\Delta} \hbox{ } v + 4 \alpha (|x|^{-1}*|v|^{2}) v 
-\frac{8}{3}\beta |v|^{\frac{2}{3}}v=\omega v \ \ \  \text{ in } \R^{3}
\end{equation}
for a suitable Lagrange multiplier $\omega\in \R$.
Moreover the corresponding time--dependent function \begin{equation}\label{solitary}
\psi(x,t)=e^{-{\bf i}\omega t}v(x)
\end{equation}
is a solution of the time-dependent Nonlinear Schr\"odinger Equation
\begin{equation}\label{RelatNLS}
{\bf i}\psi_{t}=\sqrt{1-\Delta}
\hbox{ } \psi + 4\alpha (|x|^{-1}*|\psi|^{2}) \psi-
\frac{8}{3}\beta |\psi|^{\frac{2}{3}}\psi   \ \ \     \text{ in } \R^3.
\end{equation}
As far as we know this evolutionary problem has 
not been studied
in the literature.  In this context we quote the paper
\cite{L} where it is studied the following Cauchy problem:
\begin{equation}\label{RelatNLSL}
{\bf i}\psi_{t}=  \sqrt{1-\Delta}\psi - (|x|^{-1}*|\psi|^{2}) \psi \ \ \ \text{ in } \R^3.
\end{equation}
In this case the main advantage is the smoothing effect associated 
to the Hartree nonlinearity
which allows to solve the Cauchy problem by using the classical energy estimates.
On the contrary the nonlinearity in \eqref{RelatNLS} does not enjoy the same smoothness  
and it makes more complicated the analysis of the corresponding Cauchy problem.\\
\\
{\bf Acknowledgement:} the authors would like to thank T. Cazenave, V. Georgiev 
and L.Vega for fruitful conversations.
\\
\section{Proof of Theorem \ref{bestc}}
This section is devoted to the proof
of  theorem \ref{bestc}.
\begin{prop}
There exists $C>0$ such that
$$\|\varphi\|_{L^\frac 83(\R^3)}\leq C\|\varphi\|_{\dot H^\frac 12(\R^3)}^\frac 12
\left (
\int\int_{\R^3\times \R^3}
\frac{|\varphi(x)|^2|\varphi(y)|^2}{|x-y|}dxdy\right )^\frac 18
$$
\end{prop}
{\bf Proof.} By using basic facts on Fourier transform
the previous estimate is equivalent to the following one:
\begin{equation}\label{oz}
\|\varphi\|_{L^\frac 83(\R^3)}\leq C
\|\varphi\|_{\dot H^\frac 12(\R^3)}^\frac 12
 \||\varphi|^2\|_{\dot H^{-1}(\R^3)}^\frac 14.\end{equation}
Notice that we have
the following Gagliardo-Nirenberg inequality
$$\||D|\varphi\|_{L^\frac 43(\R^3)} \leq C \|\varphi\|_{L^2(\R^3)}^\frac 13
\||D|^\frac 32 \varphi\|_{L^\frac 87(\R^3)}^\frac 23$$
that can be rewritten as follows:
$$\|\varphi\|_{L^\frac 43(\R^3)} \leq C \||D|^{-1}\varphi\|_{L^2(\R^3)}^\frac 13
\||D|^\frac 12 \varphi\|_{L^\frac 87(\R^3)}^\frac 23.$$
Next we replace $\varphi$ by $|\varphi|^2$
and we get
$$\|\varphi\|_{L^\frac 83(\R^3)}^2 \leq C \||D|^{-1}|\varphi|^2
\|_{L^2(\R^3)}^\frac 13
\||D|^\frac 12 |\varphi|^2\|_{L^\frac 87(\R^3)}^\frac 23$$
that in turn by the fractional chain rule implies
$$\|\varphi\|_{L^\frac 83(\R^3)}^2 \leq C \||D|^{-1}|\varphi|^2\|_{L^2}^\frac 13
\||D|^\frac 12 \varphi\|_{L^2(\R^3)}^\frac 23 
\|\varphi\|_{L^\frac 83(\R^3)}^\frac 23.$$
The last inequality is equivalent to \eqref{oz}.
\hfill$\Box$
\\
\\
We shall underline that new  lower bounds for the Coulomb energy involving $L^p$ spaces and homogeneous Sobolev spaces $\dot H^s$ are recently generalized in \cite{BFV}. The proof of the theorem \ref{bestc}
follows by combining the next two propositions.
In the sequel the energy $\tilde{\mathcal E}^{\alpha,\beta}$
is the one defined in \eqref{IItilde}.
\begin{prop}
The following facts are equivalent:
\begin{itemize}
 \item 
$I^{\alpha,\beta}
(\rho)=-\infty$
\item $\exists \varphi\in S(\rho) \hbox{ s.t. } 
\tilde{\mathcal E}^{\alpha,\beta}(\varphi)<0.$
\end{itemize}
\end{prop}
{\bf Proof.}
If $I^{\alpha,\beta}(\rho)=-\infty$ then there exists $\varphi\in S(\rho)$ 
such that ${\mathcal E}^{\alpha,\beta}(\varphi)<0$
and hence $\tilde{\mathcal E}^{\alpha,\beta}(\varphi)
\leq {\mathcal E}^{\alpha,\beta}(\varphi)<0$.\\
Next we prove the opposite implication.
We introduce $\varphi_\theta(x)=\theta^\frac 32 \varphi(\theta x)$
then by a scaling argument
$$\tilde {\mathcal E}^{\alpha,\beta}(\varphi_\theta)=\theta 
\tilde{\mathcal E}^{\alpha,\beta}(\varphi).
$$
Next notice that
$$\|\varphi_\theta\|_{H^\frac 12}^2-\|\varphi_\theta\|_{\dot H^\frac 12}^2
=\int \sqrt {1+|\xi|^2} \left |\hat \varphi\left (\frac \xi \theta
\right )\right |^2 \frac{d\xi}{\theta^3} -
\theta \int |\xi| |\hat \varphi(\xi)|^2 d\xi$$
$$= \int (\sqrt {1+\theta^2 |\xi|^2} - \theta|\xi|)  |\hat \varphi(\xi)|^2 d\xi 
=\int \frac 1{\sqrt {1+\theta^2 |\xi|^2} +\theta|\xi|}  |\hat \varphi(\xi)|^2 d\xi=o(1) 
\hbox{ as } \theta\rightarrow \infty.$$
Finally we get
$${\mathcal E}^{\alpha,\beta}(\varphi_\theta)
=\tilde {\mathcal E}^{\alpha,\beta}(\varphi_\theta) + \frac 12 
(\|\varphi_\theta\|_{H^\frac 12}^2-\|\varphi_\theta\|_{\dot H^\frac 12}^2)$$
$$=\theta \tilde {\mathcal E}^{\alpha,\beta}(\varphi)+o(1)$$
and hence
$$I^{\alpha,\beta}(\rho)\leq \lim_{\theta\rightarrow \infty}
{\mathcal E}^{\alpha,\beta}(\varphi_\theta)=-\infty.$$

\hfill$\Box$

\begin{prop}
The following facts are equivalent:
\begin{itemize}
\item $\tilde{\mathcal E}^{\alpha,\beta}(\varphi)\geq 0 \hbox{ } \forall \varphi\in H^\frac 12$;
\item $\left(\frac{27\alpha}{\beta^3}\right )^\frac 18\geq \sqrt 2S.$
\end{itemize}
\end{prop}

{\bf Proof.}
Let $\varphi_\theta(x)=\varphi\left (\frac x\theta\right )$
then we have 
$$\tilde{\mathcal E}^{\alpha,\beta}(\varphi)\geq 0 
\hbox{ } \forall \varphi\in H^\frac 12 $$
if and only if
$$\tilde{\mathcal E}^{\alpha,\beta}(\varphi_\theta)\geq 0
\hbox{ } \forall \varphi\in H^\frac 12, \theta\in (0,\infty).$$
By explicit computation this is equivalent to
$$\frac 12 \theta^{2} \|\varphi\|_{\dot H^\frac 12}^2
+ \alpha \theta^{5}
\int \int \frac{|\varphi(x)|^2 |\varphi(y)|^2}
{|x-y|} dxdy-\beta \theta^{3} \|\varphi\|_{\frac 83}^\frac 83\geq 0
$$$$\hbox{ } \forall \varphi\in H^\frac 12, \theta\in (0,\infty).$$
Hence the condition $\tilde{\mathcal E}^{\alpha,\beta}(\varphi)\geq 0 
\hbox{ } \forall \varphi\in H^\frac 12 $ 
can be rewritten as follows:
\begin{equation}\label{ozaw}
\inf_{\theta \in (0,\infty)}\psi^{\alpha,\beta}_\varphi (\theta)\geq 0
\hbox{ } \forall \varphi\in H^\frac 12, \theta\in (0,\infty)
\end{equation}
where
$$\psi^{\alpha,\beta}_\varphi (\theta)=
\frac 12 \|\varphi\|_{\dot H^\frac 12}^2
+ \alpha \theta^{3}
\int \int \frac{|\varphi(x)|^2 |\varphi(y)|^2}
{|x-y|} dxdy-\beta \theta \|\varphi\|_{\frac 83}^\frac 83\geq 0.
$$
By elementary computation we get
$$\inf_{(0,\infty)} \psi^{\alpha,\beta}_\varphi (\theta)
= \psi^{\alpha,\beta}_\varphi \left( \|\varphi\|_\frac 83^\frac 43 \sqrt 
\frac{\beta (\int \int \frac{|\varphi(x)|^2 |\varphi(y)|^2}
{|x-y|} dxdy)^{-1}}
{3\alpha} \right)$$$$
=\frac 12 \|\varphi\|_{\dot H^\frac 12}^2 
+\left( \alpha \left(\frac{\beta}{3\alpha} \right)^\frac 32 
- \beta \sqrt{\frac{\beta}{3\alpha}}\right) \frac{\|\varphi\|_\frac 83^4}
{\sqrt{\int \int \frac{|\varphi(x)|^2 |\varphi(y)|^2}
{|x-y|} dxdy} } 
$$
$$=\frac 12 \|\varphi\|_{\dot H^\frac 12}^2 -\frac 23 \beta \sqrt 
{\frac \beta{3\alpha}}
\frac{\|\varphi\|_\frac 83^4}
{\sqrt{\int \int \frac{|\varphi(x)|^2 |\varphi(y)|^2}
{|x-y|} dxdy} }.$$
Hence the condition \eqref{ozaw}
becomes 
$$4 \sqrt{\frac{\beta^3}{27\alpha}} \|\varphi\|_\frac 83^4
\leq \|\varphi\|_{\dot H^\frac 12}^2 \sqrt{\int \int \frac{|\varphi(x)|^2 |\varphi(y)|^2}
{|x-y|} dxdy} \hbox{ } \forall \varphi\in \dot H^\frac 12$$
and we can conclude since by definition $S$ is the best constant
in the inequality
$$\|\varphi\|_\frac 83
\leq S \|\varphi\|_{\dot H^\frac 12}^\frac 12  \left( \int \int \frac{|\varphi(x)|^2 |\varphi(y)|^2}
{|x-y|} dxdy\right )^\frac 18  \hbox{ } \forall \varphi\in \dot H^\frac 12.$$

\hfill$\Box$

\section{Proof of Theorem \ref{XXXX}}
First we quote a recent result to avoid vanishing in $\dot H^{s}$. It is a generalization of the Lieb Translation Lemma which
holds in $H^1$, see \cite{LL}.
\begin{lem}[Lieb Translation Lemma in $\dot H^{s}$, \cite{BFV}]\label{LiebIntro}
Let $s>0$, $1<p<\infty$ and $u_n\in \dot H^s(\R^d)\cap L^p(\R^d)$ be a sequence with
\begin{equation}\label{hyp1}
\sup_n \left( \|u_n\|_{\dot H^{s}} +\|u_n\|_{L^p} \right) <\infty
\end{equation}
and, for some $\eta>0$, (with $|\cdot |$ denoting Lebesgue measure)
\begin{equation}\label{hyp2}
\inf_n \left|\{ |u_n|>\eta \}\right|>0 \,.
\end{equation}
Then there is a sequence $(x_n)\subset\R^d$ such that a subsequence of $u_n(\cdot+ x_n)$ has a weak limit $u\not\equiv 0$ in $\dot H^s(\R^d)\cap L^p(\R^d)$.
\end{lem}

Now we state four  propositions that are important for the sequel.

\begin{prop}\label{facile83}
For every $\alpha, \beta>0$ there exists $\rho_0=\rho_0(\beta)>0$ such that
$$I^{\alpha,\beta}(\rho)\geq 0 \hbox{ } \ \ \  \forall \ 0<\rho<\rho_0.$$
Moreover if $u_{n}\in S(\rho)$
is a minimizing sequence for $I^{\alpha,\beta}(\rho)$ with $0<\rho<\rho_0$
then $$\sup_n \|u_{n}\|_{H^\frac 12}<\infty.$$
\end{prop}
{\bf Proof.}
It follows from the following estimate
$${\mathcal E}^{\alpha,\beta}(\varphi)\geq \frac 12 \|\varphi\|_{H^\frac 12}^2
- C\|\varphi\|_{2}^\frac 23 \|\varphi\|_{H^\frac 12}^2
$$
where we have used the H\"older inequality in conjunction
with the Sobolev embedding $H^\frac 12\subset L^3$.

\hfill$\Box$

\begin{prop}\label{cont83}
Let $\alpha, \beta>0$ be fixed and $\bar \rho=\bar \rho(\alpha, \beta)>0$ 
be such that $I^{\alpha,\beta} (\rho)>-\infty$ for $\rho\in (0, \bar \rho)$.
Then the function
$$(0,\bar \rho)\ni \rho\rightarrow I^{\alpha,\beta}(\rho)\in \R$$
is continuous.
\end{prop}
{\bf Proof.} 
Assume it is not continuous, then there exists a sequence $\rho_n$ and $\epsilon>0$ such that
$\lim_{n\rightarrow} \rho_n=\bar \rho>0$ and
$|I^{\alpha,\beta}(\rho_n)-I^{\alpha,\beta}(\bar \rho)|\geq \epsilon>0.$ 
In particular up to subsequence
we can assume that
either
\begin{equation}\label{i}
I^{\alpha,\beta}(\rho_n)-I^{\alpha,\beta}(\bar \rho)\geq \epsilon
\end{equation} or 
\begin{equation}\label{ii}
I^{\alpha,\beta}(\bar \rho)-I^{\alpha,\beta}(\rho_n)\geq \epsilon.
\end{equation}
First we shall prove that \eqref{i} cannot occur.
We fix $w\in H^\frac 12$ such that 
\begin{equation}\label{w}
w\in S(\bar \rho) \hbox{ and }{\mathcal E}^{\alpha,\beta}(w)-
I^{\alpha,\beta}(\bar \rho)\leq \frac \epsilon 2
\end{equation}
and we introduce 
$$w_n=\sqrt{\frac{\rho_n}{\bar \rho}} w.$$ 
Notice that
\begin{equation}\label{levpar}
w_n\in S(\rho_n) \hbox{ and } \lim_{n\rightarrow \infty} {\mathcal E}^{\alpha,\beta}(w_n)
={\mathcal E}^{\alpha,\beta}(w).
\end{equation}
By combining \eqref{levpar} with \eqref{w} we get the existence of
$\bar n\in \N$ such that 
\begin{equation}\label{penul}
I^{\alpha,\beta}(\rho_n)\leq {\mathcal E}^{\alpha,\beta}(w_n)
\leq {\mathcal E}^{\alpha,\beta}(w)+\frac \epsilon 4
\leq  I^{\alpha,\beta}(\bar \rho) +\frac 34 \epsilon \hbox{ } \forall n>\bar n
\end{equation}
which is in contradiction with \eqref{i}.\\
In order to contradict \eqref{ii} we argue as follows.
Let $v_n\in H^\frac 12$ 
such that
\begin{equation}\label{vn}
v_n\in S(\rho_n) \hbox{ and }{\mathcal E}^{\alpha,\beta}(v_n)-I^{\alpha,\beta}(\rho_n)\leq \frac \epsilon 2.
\end{equation}
We  state the following\\
\\
{\bf Claim} {\em We can choose a sequence $v_n$ that satisfies \eqref{vn}
and moreover $$\sup_n \|v_n\|_{H^\frac 12}<\infty.$$}
\\
\\
By assuming the claim it is easy to prove that
\begin{equation}\label{ivs}\lim_{n\rightarrow \infty} ({\mathcal E}^{\alpha,\beta}(v_n)- 
{\mathcal E}^{\alpha,\beta}(u_n))=0.
\end{equation}
where
$$u_n=\sqrt{\frac{\bar \rho}{\rho_n}} v_n\in S(\bar \rho).$$
By combining \eqref{vn} with \eqref{ivs} we get the existence of $\bar n\in \N$ such that
$$I^{\alpha,\beta}(\bar \rho)
\leq {\mathcal E}^{\alpha,\beta}(u_n)\leq {\mathcal E}^{\alpha,\beta}(v_n)
+\frac \epsilon 4 \leq I^{\alpha,\beta}(\rho_n)
+\frac 34 \epsilon \hbox{ } \forall n>\bar n$$
hence contradicting \eqref{ii}.\\
Next we shall prove the claim. Notice that if \eqref{ii} is true then 
$$K=\sup_n I^{\alpha,\beta}(\rho_n)<\infty$$ and  we deduce that $v_n$ can be choosen in such a way that:
$$ K +1 \geq  {\mathcal E}^{\alpha,\beta}(v_{n}) \geq h_{\rho_n}(\|v_{n}\|_{H^\frac 12})$$
where $h_{\rho_n}(t)=\frac 12 t^2 - C \rho_n^{\frac{1}{3}}t^{2}$.
It is now easy to deduce the claim since for every
$M>0$ there exists $R>0$ such that
$$h_{\rho_n}(t)\geq M \hbox{ } \forall t\geq R \hbox{ } \forall n\in \N.$$

\hfill$\Box$

\begin{prop}\label{cruccont}
For every $\alpha, \beta>0$ there exists $\rho_1=\rho_1(\alpha, \beta)>0$ such that
\begin{equation}\label{hyp3383}
\frac{I^{\alpha,\beta}(\rho)}{\rho}<\frac{1}{2} \hbox{ } \ \ \forall \ 0<\rho<\rho_1.
\end{equation}
Moreover 
\begin{equation}\label{hyp2283}
\lim_{\rho \rightarrow 0}\frac{I^{\alpha,\beta}(\rho)}{\rho}=\frac{1}{2}.
\end{equation}
\end{prop}
{\em Proof of \eqref{hyp3383}.}
\\
\\
We introduce the functional
\begin{equation}\label{fstorto}
{\mathcal F}^{\alpha,\beta}(u)
={\mathcal E}^{\alpha,\beta} (u)-\frac 12 \|u\|_2^2={\mathcal E}^{\alpha,\beta} (u)
-\frac 12 \|\hat u\|_2^2\end{equation}
$$=\frac 12 \int\frac{|\xi|^2}{1+\langle \xi \rangle} |\hat u|^2 d\xi
+\alpha \int \int \frac{|u(x)|^2|u(y)|^2}{|x-y|}dxdy
-\beta \|u\|_{\frac{8}{3}}^{\frac{8}{3}}$$
where we have used the Plancharel identity.
Notice that \eqref{hyp3383} is equivalent to show that
\begin{equation}\label{claimequiv}
 \inf_{u\in S(\rho)} {\mathcal F}^{\alpha,\beta}(u)<0 \hbox{ } 
\forall 0<\rho<\rho_1\end{equation}
with $\rho_1$ small enough.
In order to prove \eqref{claimequiv} we fix $\varphi\in C^\infty_0(\R^3)$ such that $\varphi \in S(1)$
and and we introduce $\varphi_\theta=\theta^\gamma \varphi(\theta x)$
where $\gamma$ will be choosen later.
Notice that
by looking at the expression of ${\mathcal F}^{\alpha,\beta}$ in \eqref{fstorto} we get
$$\inf_{u\in S(\theta^{2\gamma-3})} 
{\mathcal F}^{\alpha,\beta}(u)
\leq {\mathcal F}^{\alpha,\beta} (\varphi_\theta)\leq 
\frac 12 \int |\xi|^2|\hat \varphi_\theta|^2 d\xi 
$$$$+  
\alpha \int \int \frac{|\varphi_\theta(x)|^2|\varphi_\theta(y)|^2}{|x-y|}dxdy-\beta 
\|\varphi_\theta\|_{\frac{8}{3}}^{\frac{8}{3}}
$$
$$=\frac 12 \theta^{2\gamma-1}\|\varphi\|_{\dot H^1}^2 + \alpha \theta^{4\gamma-5} 
\int \int \frac{|\varphi(x)|^2|\varphi(y)|^2}{|x-y|}dxdy-\beta \theta^{\frac{8}{3}\gamma-3}
\|\varphi\|_{\frac{8}{3}}^{\frac{8}{3}}.$$
Notice that the r.h.s. above is negative
for $0<\gamma<\bar \gamma$
provided that we can choose $\gamma$ such that
$$2\gamma-1>0, 4\gamma-5>0,\frac{8}{3}\gamma-3>0$$ 
$$\frac{8}{3}\gamma-3< 2\gamma-1, \frac{8}{3}\gamma-3<4\gamma-5.$$
In fact the conditions above are satisfied for any $\gamma \in (\frac{3}{2},3)$.
Notice that $\varphi_\theta \in S(\theta^{2\gamma-3})$, with  $\theta^{2\gamma-3}\rightarrow 0$  when $\theta \rightarrow 0$  if $\gamma>\frac{3}{2}$.\\

{\em Proof of \eqref{hyp2283}.}

Due to \eqref{hyp3383} it is sufficient to prove that 
$$\liminf_{\rho\rightarrow 0} \frac{I^{\alpha,\beta}(\rho)}{\rho}\geq \frac 12.$$
For every $\rho>0$ we fix a minimizing sequence $u_{n}\in S(\rho)$ for $I^{\alpha,\beta}(\rho)$
hence we have
$$
\frac{I^{\alpha,\beta}(\rho)}{\rho}\geq \limsup_{n\rightarrow \infty} 
\left (\frac 12 \rho^{-1}\|u_{n}\|_{H^\frac 12}^2
- \beta \rho^{-1}\|u_{n}\|_{\frac{8}{3}}^{\frac{8}{3}}\right ).$$
Notice that
$\limsup_{n\rightarrow \infty} \frac 12 \rho^{-1}\|u_{n}\|_{H^\frac 12}^2\geq
\limsup_{n\rightarrow \infty} \frac 12 \rho^{-1}\|u_{n}\|_{L^2}^2=\frac 12$
hence it is sufficient to prove that
$$\limsup_{\rho\rightarrow 0} (\limsup_{n\rightarrow \infty} \rho^{-1}\|u_{n}\|_{\frac{8}{3}}^{\frac{8}{3}})=0.$$
This fact will follow by combining next 
claim with the usual Sobolev embedding $H^\frac 12\subset L^{\frac{8}{3}}$.  
\\
{\bf Claim}
\begin{equation}\label{jacca}
\exists \bar \rho>0, C>0 \hbox{ s.t. } 
\limsup_{n\rightarrow \infty} \|u_{n}\|_{H^\frac 12}<C\sqrt \rho \hbox{ } 
\forall \rho<\bar \rho.
\end{equation}
By combining the H\"older inequality and the Sobolev embedding and 
\eqref{hyp3383} we get:
\begin{equation}\label{hdelta}\frac 12 \rho> I^{\alpha,\beta}(\rho)=\lim_{n\rightarrow \infty} 
{\mathcal E}^{\alpha,\beta}(u_{n})\geq
\limsup_{n\rightarrow \infty} h_{\rho}(\|u_{n}\|_{H^\frac 12}) \hbox{ } 
\forall 0<\rho<\rho_1
\end{equation}
where $h_{\rho}(t)=\frac 12 t^2 - C \rho^{\frac{1}{3}}t^{2}\geq \frac 14 t^2$ (for $\rho$ 
small enough) 
and $u_{n}\in S(\rho)$
is a minimizing sequence for $I^{\alpha,\beta}(\rho)$. This implies that 
$$\limsup_{n\rightarrow \infty} \|u_{n}\|_{H^\frac 12}\leq C \sqrt \rho$$
for $0<\rho<\rho_1$
with $\rho_1$ suitable small number. 

\hfill$\Box$

\begin{prop}\label{ccom}[Concentration-Compactness]
Let $\alpha, \beta>0$ be fixed. Let $\rho>0$ be such that
$I^{\alpha,\beta}(\rho)>-\infty$. Assume moreover
\begin{equation}\label{subadd}
\rho I^{\alpha,\beta}(\rho')< \rho' 
I^{\alpha,\beta}(\rho) \hbox{ } \forall \hbox{ } 0<\rho'<\rho.
\end{equation}
Then for every minimizing sequence 
$u_{n}\in S(\rho)$ for $I^{\alpha,\beta}(\rho)$ 
there exists, up to subsequence, $\tau_n \in \R^3$  such that
$u_{n}(.+\tau_n)$ converge strongly to $\bar u$ in $H^\frac 12$.
\end{prop}

{\bf Proof of Proposition \ref{ccom}.}
\\
Recall that by  Proposition \ref{facile83} 
we can assume $\sup_n \|u_{n}\|_{H^\frac 12}<\infty$.\\
\\
{\em First step: no-vanishing}
\\
\\
First we prove the following
\\
\\
{\bf Claim} $\exists \epsilon_0>0 \hbox{ s.t. } \|u_{n}\|_{\frac{8}{3}}\geq \epsilon_0$
\\
\\
Assume it is not true then 
$\lim_{n\rightarrow \infty} \|u_{n}\|_{\frac{8}{3}}=0$ and in particular
$$I^{\alpha,\beta}(\rho)=\lim_{n\rightarrow \infty} \frac 12 \|u_{n}\|_{H^\frac 12}^2 
$$$$+ \alpha \int\int \frac{|u_{n}(x)|^2 |u_{n}(y)|^2}{|x-y|}dxdy 
- \beta\|u_{n}\|_{\frac{8}{3}}^{\frac{8}{3}}\geq  \lim_{n\rightarrow \infty} \frac 12 \|u_{n}\|_{2}^2 
=\frac 12 \rho$$
which is in contradiction with  \eqref{hyp3383}.\\
By combining the claim with the fact that $||u_n||_{L^2}<\infty, ||u_n||_{L^3}<\infty$, by the well known \emph{PQR} Lemma, see \cite{FLL}, one gets the existence 
of $\eta>0$ such that
$$\inf_n \left|\{ |u_n|>\eta \}\right|>0.$$
By Lieb Translation Lemma \ref{LiebIntro} in $\dot H^s$, $s>0$, see \cite{BFV},  we get
the existence, up to subsequence, of 
$\tau_n$ such that
$$v_{n}=u_{n}(.+\tau_n)$$ has a weak limit $\bar v$ different from zero.
\\
{\em Second step: $v_{n}$ converges strongly to $\bar v$ in $H^\frac 12$}
\\
\\
It is sufficient to prove that
$v_{n}$ converges strongly to $\bar v$ in $L^2$
(then the strong convergence in $H^\frac 12$ follows 
by the fact that $v_{n}$ is a minimizing sequence for $I^{\alpha,\beta}(\rho)$).
In particular it is sufficient to prove that $\|\bar v\|_2^2=\rho$.
Assume by the absurd that $\|\bar v\|_2^2=\delta\in (0,\rho)$,
then since $L^2$ and $H^\frac 12$ are Hilbert spaces we get:
\begin{equation}\label{orthL2}\|v_{n}-\bar v\|_2^2=\rho - \delta + o(1)\end{equation}
and also
\begin{equation}\label{orthH12}\|v_{n}-\bar v\|_{H^\frac 12}^2=
\|v_{n}\|_{H^\frac 12}^2 - \|\bar v\|_{H^\frac 12}^2 + o(1).\end{equation}
Moreover, up to subsequence, we can assume that
$$v_{n}(x)\rightarrow \bar v(x) \hbox{ a.e. }x\in \R^3.$$
Hence via the Br\'ezis-Lieb Lemma (see \cite{BL})
we get
\begin{equation}\label{brelifr}\|v_{n}-\bar v\|_{p}^p=
\|v_{n}\|_{p}^p - \|\bar v\|_{p}^p + o(1)
\end{equation}
and by \cite{BFV}
\begin{equation}\label{brelinonlocal}
\int\int \frac{|(v_{n}-\bar v)(x)|^2 |(v_{n}-\bar v)(y)|^2}{|x-y|} dxdy
\end{equation}
$$=\int\int \frac{|(v_{n}(x)|^2 |(v_{n}(y)|^2}{|x-y|} dxdy
-\int\int \frac{|\bar v(x)|^2 |(\bar v)(y)|^2}{|x-y|} dxdy+o(1).$$
By combining \eqref{orthL2}, \eqref{orthH12}, \eqref{brelifr}, \eqref{brelinonlocal} and
the fact that $v_{n}$ is a minimizing sequence for
$I^{\alpha,\beta}(\rho)$ we get
$$I^{\alpha,\beta}(\rho)=
{\mathcal E}^{\alpha,\beta}(v_{n})+o(1) = {\mathcal E}^{\alpha,\beta}
(v_{n}-\bar v) + {\mathcal E}^{\alpha,\beta}(\bar v)+ o(1)$$
$$\geq I^{\alpha,\beta}(\rho-\delta + o(1))
+ I^{\alpha,\beta}(\delta)+o(1)$$
and in particular by the continuity of the function $I^{\alpha,\beta}(\rho)$ 
(see proposition \ref{cont83}) we get
\begin{equation}\label{abssubbad}
I^{\alpha,\beta}(\rho)\geq I^{\alpha,\beta}(\rho-\delta)
+ I^{\alpha,\beta}(\delta).\end{equation}
Next notice that by \eqref{subadd}
we get 
$$I^{\alpha,\beta}(\rho-\delta)>\frac{\rho-\delta}{\rho} I^{\alpha,\beta}(\rho) 
\hbox{ and } I^{\alpha,\beta}(\delta)>
\frac{\delta}{\rho} I_p^{\alpha,\beta}(\rho)$$
which imply
$$I^{\alpha,\beta}(\rho-\delta)+ I^{\alpha,\beta}(\delta)> I^{\alpha,\beta}(\rho)$$
hence contradicting \eqref{abssubbad}.

\hfill$\Box$

{\bf Proof of Theorem  \ref{XXXX}.}

First we prove the existence of a sequence of ground states for $I^{\alpha,\beta}(\rho_n)$
\\
\\
{\bf Claim} $\exists$ a sequence $\rho_n\rightarrow0$, and $u_n \in S(\rho_n)$  \hbox{ s.t. } $I^{\alpha,\beta}(\rho_n)= {\mathcal E}^{\alpha,\beta}(u_n).$\\
\\
The proof of the claim  follows from a continuity argument.  Fix $\epsilon>0$, and define
$$\rho_{\epsilon}:= \inf \{\rho>0, \text{ s.t } \frac{I^{\alpha,\beta}(\rho)}{\rho}=\frac{1}{2}-\epsilon \}.$$
By Proposition \ref{cruccont} and Proposition \ref{cont83}, $\rho_{\epsilon}>0$ and 
\begin{equation}\label{subadd1}
\frac{I^{\alpha,\beta}(\rho_{\epsilon})}{\rho_{\epsilon}}<\frac{I^{\alpha,\beta}(\rho)}{\rho} \ \ \  \forall 0<\rho<\rho_{\epsilon}.
\end{equation}
The existence of a ground state for  $I^{\alpha,\beta}(\rho_{\epsilon})$ follows from Proposition \ref{ccom}  observing that \eqref{subadd1} is exactly condition \eqref{subadd}.
Sending $\epsilon\rightarrow 0$ we get the claim.\\
\\
Now we shall prove the existence of ground states for all $0<\rho<\bar \rho(\alpha, \beta).$ By Proposition \ref{ccom} it is sufficient to prove the monotonicity
of $\frac{I^{\alpha,\beta}(\rho)}{\rho}$ for all $0<\rho<\bar \rho(\alpha, \beta).$  Fix $\rho>0$, and define $c=\min_{(0, \rho]} \frac{I^{\alpha,\beta}(s)}{s}<\frac{1}{2}$ and 
$$\rho_0:= \inf \{s \in (0, \rho] , \text{ s.t } \frac{I^{\alpha,\beta}(s)}{s}=c \}.$$ 
We have to prove that $\rho_0=\rho.$  \\
Assume by contradiction that $\rho_0<\rho$. Following  the claim let us call $u_{\rho_0}$ the ground state for $I^{\alpha,\beta}(\rho_0)$.
If we assume that monotonicity $\frac{I^{\alpha,\beta}(s)}{s}$
breaks at $\rho_0$ hence the following shall hold
\begin{equation}\label{finarg}
\frac{{\mathcal E}^{\alpha,\beta}(u_{\rho_0})}{\rho_0}=\frac{I^{\alpha,\beta}(\rho_0)}{\rho_0}\leq \frac{I^{\alpha,\beta}(\theta^2 \rho_0)}{\theta^2 \rho_0}\leq \frac{{\mathcal E}^{\alpha,\beta}(\theta u_{\rho_0})}{\theta^2 \rho_0} 
\end{equation}
for all  $0<\theta<1$ and for a sequence $\theta_n>1$ with $\lim_{n \rightarrow \infty}\theta_n=1$ (we shall consider only a sequence  because $\frac{I^{\alpha,\beta}(\rho)}{\rho}$ can be fact oscillating for $\rho>\rho_0$).
Inequality \eqref{finarg} implies that
$$\frac{d}{d \theta}\left( \theta^2 {\mathcal E}^{\alpha,\beta}(u_{\rho_0}) -{\mathcal E}^{\alpha,\beta}( \theta u_{\rho_0})\right)_{\theta=1}=0$$
which is equivalent to
\begin{equation}\label{class283}
2\alpha \int \int \frac{|u_{\rho_0}(x)|^2|u_{\rho_0}(y)|^2}{|x-y|}dxdy
-\frac 23\beta \|u_{\rho_0}\|_\frac 83^\frac 83=0.
\end{equation}
To conclude the proof it suffices to apply  Hardy-Littlewood-Sobolev inequality
and the interpolation inequality to get
$$\|u_{\rho_0}\|_\frac 83^\frac 83=3\frac{\alpha}{\beta} \int \int \frac{|u_{\rho_0}(x)|^2|u_{\rho_0}(y)|^2}{|x-y|}dxdy
\leq C \|u_{\rho_0}\|_{\frac{12}5}^4\leq C \rho_0^\frac 23 \|u_{\rho_0}\|_{\frac 83}^\frac 83$$
which cannot hold if $\rho_0$ and hence $\rho$ is sufficient small.
\hfill$\Box$
\section{Proof of Theorem \ref{homogeneous}}\label{sec6}

We shall need the following lemma.
\begin{lem}\label{alt}
The following dichotomy happens:
\begin{equation}\label{alter}
\hbox{ either } \tilde I^{\alpha,\beta}(\rho)=0 \hbox{ or } 
I^{\alpha,\beta}(\rho)=-\infty.
\end{equation}
Moreover there exists $\tilde \rho>0$ such that
\begin{equation}\label{piccolirho}
\tilde I^{\alpha,\beta}(\rho)=0 \hbox{  } \forall \rho\in (0, \tilde \rho)
\end{equation}
\end{lem}
{\bf Proof.}
{\em First step: $\tilde I^{\alpha,\beta}(\rho)\leq 0$}
\\
\\
We fix $\varphi\in C^\infty_0(\R^3)$ such that $\|\varphi\|_2^2=\rho$ and 
$\varphi_\theta=\theta^\frac 32 \varphi(\theta x)$.
Then $\|\varphi\|_2^2=\rho$.\\
By direct computation
\begin{itemize}
\item $\|\varphi_\theta\|_\frac 83^\frac 83 = \theta \|\varphi\|_\frac 83^\frac 83$
\item $\int\int 
\frac{|\varphi_\theta(x)|^2|\varphi_\rho(y)|^2}{|x-y|}dxdy=\theta 
\int\int 
\frac{|\varphi(x)|^2|\varphi(y)|^2}{|x-y|}dxdy.$
\item $\|\varphi_\theta\|_{\dot H^\frac 12}^2=\theta\|\varphi\|_{\dot H^\frac 12}^2$
\end{itemize}
In particular we get
$$\tilde {\mathcal E}^{\alpha,\beta}
(\varphi_\theta)=\theta \tilde {\mathcal E}^{\alpha,\beta}\varphi)$$
which implies
$$\tilde I^{\alpha,\beta}
(\rho)\leq \lim_{\theta\rightarrow 0} \tilde {\mathcal E}^{\alpha,\beta}(\varphi_\theta)=0.$$
\\
{\em Second step: if $\tilde I^{\alpha,\beta}(\rho)<0$ then 
$\tilde I^{\alpha,\beta}(\rho)=-\infty$}
\\
\\
Let $\varphi\in S(\rho)$ be such that $\tilde {\mathcal E}^{\alpha,\beta}(\varphi)<0$
then arguing as above
we get 
$$\tilde {\mathcal E}^{\alpha,\beta}
(\varphi_\theta)=\theta \tilde {\mathcal E}^{\alpha,\beta}(\varphi)$$
where
$\varphi_\theta=\theta^\frac 32 \varphi(\theta x)$.
Hence 
$$\tilde I^{\alpha,\beta} (\rho)\leq \lim_{\theta\rightarrow \infty} 
\tilde {\mathcal E}^{\alpha,\beta}(\varphi_\theta)=
\lim_{\theta\rightarrow \infty} \theta 
\tilde {\mathcal E}^{\alpha,\beta}(\varphi)=-\infty.$$
The proof of \eqref{alter} follows easily.\\
Next we focus on \eqref{piccolirho}.
Notice that by combining H\"older inequality with
the Sobolev embedding $\dot H^\frac 12\subset L^3$ we get
\begin{equation}\label{sobpiuh}
\tilde {\mathcal E}^{\alpha,\beta}(u)\geq \frac 12 \|u\|_{\dot H^\frac 12}
-\beta \|u\|_3^2\|u\|_2^\frac 23\geq \
\frac 12 \|u\|_{\dot H^\frac 12}
-C \|u\|_{\dot H^\frac 12}^2 \rho^\frac 13 \hbox{ } \forall u\in S(\rho).
\end{equation}
In particular if $\rho$ is small then 
$\tilde {\mathcal E}^{\alpha,\beta}(u)\geq 0$ for any $u\in S(\rho)$ and hence
$$\tilde I^{\alpha,\beta}(\rho)\geq 0.$$
By combining this fact with \eqref{alter}
we deduce \eqref{piccolirho}.

\hfill$\Box$

{\bf Proof of theorem \ref{homogeneous}.}
Let $\rho_*>0$ be such that
$$\frac 12- C \rho_*^\frac 13>0$$
where $C$ is the universal constant that appears in \eqref{sobpiuh}.
Let $\tilde \rho$ be as in lemma \ref{alt}. Then by using lemma \ref{alt}
$\tilde I^{\alpha, \beta}(\rho)=0$ for every $\rho<min\{\tilde \rho, \rho_*\}$.  
By combining this fact with \eqref{sobpiuh} we deduce that if $u_n$ is a minimizing sequence
for $\tilde I^{\alpha,\beta}(\rho)$ with $\rho<min\{\tilde \rho, \rho_*\}$
then 
$$\lim_{n\rightarrow 0}\|u_n\|_{\dot H^\frac 12}=0$$
In particular it implies that if $v\in S(\rho)$ is a minimizer for $\tilde I^{\alpha,\beta}(\rho)$
with $\rho<min\{\tilde \rho, \rho_*\}$
then $v=0$ (which is absurd since if $v\in S(\rho)$ for $\rho>0$ then $v\neq 0$).

\hfill$\Box$


\begin{thebibliography}{}

\bibitem{BFV} {J.Bellazzini, R.L. Frank, N. Visciglia, }{\sl Maximizers for Gagliardo-Nirenberg inequalities and related non-local problems },  arXiv:1308.5612 (2013)
\bibitem{BS}{J. Bellazzini and G. Siciliano, }{\sl Stable standing waves for a class of nonlinear Schr\"odinger-Poisson equations, }Z. Angew. Math. Phys., vol. 62, (2011), n. 2, pp. 267--280.
\bibitem{BS2}{J. Bellazzini and G. Siciliano, }{\sl Scaling properties of functionals and existence of constrained minimizers}, J. Funct. Analysis, vol. 261, (2011), n. 9,
pp. 2486-2507.
\bibitem{BL} {H. Brezis, E. Lieb, }{\sl A relation between pointwise convergence of functions and convergence of functionals, } { Proc. Amer. Math. Soc.  88  (1983), no. 3, 486--490.}
\bibitem{CL} {I. Catto, P.L. Lions,} {\sl Binding of atoms and stability of molecules in Hartree and Thomas-Fermi type theories. I. A necessary and sufficient condition for the stability of general molecular systems,} Comm. Partial Differential Equations 17 (1992), no. 7-8, 1051–1110
\bibitem{CDSS}{I. Catto, J. Dolbeaut, O. Sanchez, J. Soler, }{Existence of steady states for the Maxwell-Schršdinger-Poisson system: exploring the applicability of the concentration-compactness principle},  Math. Models Methods Appl. Sci. 23 (2013), no. 10, 1915Ð1938.
\bibitem{L}{E. Lenzmann},{\sl Well-posedness for semi-relativistic Hartree equations of critical type,}  Math. Phys. Anal. Geom. 10 (2007), no. 1, 43–64
\bibitem{LL} {E. Lenzmann, M. Lewin,} {\sl Minimizers for the Hartree-Fock-Bogoliubov theory of neutron stars and white dwarfs,} Duke Math. J. 152 (2010), no. 2, 257–315
\bibitem{LLo} {E.Lieb, M. Loss,} {\sl Analysis}, American Mathematical Society, Providence R.I. (1997)
\bibitem{LY} {E. Lieb, H. T. Yau,} {\sl The Chandrasekhar theory of stellar collapse as the limit of quantum mechanics, } Comm. Math. Phys. 112 (1987), no. 1, 147–174
\bibitem{FJL} {J. Frohlich, B. L. G. Jonsson, E. Lenzmann,} {\sl Boson stars as solitary waves.} Comm. Math. Phys. 274 (2007), no. 1, 1–30.
\bibitem{FLL} {J. Fr\"ohlich,  E. H. Lieb, M. Loss} , \emph{Stability 
of Coulomb systems with magnetic fields. I. The one-electron atom.},
Comm. Math. Phys. 104 (1986), no. 2, 251--270.
\bibitem{L} {P. L. Lions, }{\sl The concentration-compactness principle in the Calculus of Variation. The locally compact case,
part I and II, }{ Ann. Inst. H. Poincare Anal. Non Lineaire 1 (1984), 109--145 and 223--283.}
\bibitem{SS} O. Sanchez, J. Soler, {\sl Long time dynamics of the  Schr\"odinger-Poisson-Slater system}, {Journal of Statistical Physics 114 (2004), 179--204.}
\bibitem{Sl} {J.C. Slater,} {\sl A simplification of the Hartree-Fock method,} Phys. Rev. 81 (1951), 385-390
\end{thebibliography}
\end{document}